\documentstyle[12pt,epsf]{article}
\newlength{\abstractwidth}
\renewcommand{\thefootnote}{\fnsymbol{footnote}}
\renewcommand{\thanks}[1]{\footnote{#1}} 
\newcommand{\starttext}{
\setcounter{footnote}{0}
\renewcommand{\thefootnote}{\arabic{footnote}}}
\renewcommand{\theequation}{\thesection.\arabic{equation}}
\newcommand{\be}{\begin{equation}}
\newcommand{\bea}{\begin{eqnarray}}
\newcommand{\eea}{\end{eqnarray}}
\newcommand{\beq}{\begin{equation}}
\newcommand{\ee}{\end{equation}}
\newcommand{\eeq}{\end{equation}}

\def\ba{\begin{eqnarray}}
\def\ea{\end{eqnarray}}

\def\12{{1 \over 2}}
\def\32{{3 \over 2}}
\def\72{{7 \over 2}}
\def\92{{9 \over 2}}

\begin{document}
\renewcommand{\theequation}{\thesection.\arabic{equation}}
\begin{titlepage}
\bigskip
\rightline{SU-ITP 00-14} \rightline{hep-th/0011164}

\bigskip\bigskip\bigskip\bigskip

\centerline{\Large \bf {Comments Concerning the CFT Description of
}} \centerline{\Large \bf { Small Objects in AdS}}

\bigskip\bigskip
\bigskip\bigskip

\medskip
 \centerline{\it
V. Hubeny $^1$, L. Susskind $^2$ and N. Toumbas $^2$ }
\bigskip

\centerline{$^1$Department of Physics}\centerline{University of
California}
\centerline{Santa Barbara, CA 93106}

\medskip

\centerline{$^2$Department of Physics} \centerline{Stanford
University} \centerline{Stanford, CA 94305-4060}
\bigskip\bigskip
\begin{abstract}
In this paper we resolve a contradiction posed in a recent paper
by Horowitz and Hubeny. The contradiction concerns the way small
objects in AdS space are described in the holographic dual CFT
description.

\medskip
\noindent
\end{abstract}

\end{titlepage}
\starttext \baselineskip=18pt \setcounter{footnote}{0}

\setcounter{equation}{0}
\section{The Apparent Contradiction }
According to the Holographic Principle
\cite{'thooft}\cite{susskind} objects deep in the interior of a
spatial region should have a description in terms of a holographic
theory that in some sense lives on the region's boundary. A
concrete realization of this idea has been given by the AdS/CFT
duality \cite{malda}\cite{GKP}\cite{witten}. It has become a
subject of active investigation to find out exactly how particular
objects in $AdS$ space are represented in the corresponding
conformal field theory.

There are two apparently contradictory claims in the literature
concerning this question. According to \cite{pst}\cite{precursors}
the field theoretic representation of an object or event far from
the $AdS$ boundary is through  nonlocal operators such as Wilson
loops whose degree of nonlocality increases as the object recedes
from the boundary. This follows from the usual UV/IR
correspondence \cite{susskind1}.

In apparent contradiction with this view, Horowitz and Hubeny in
an extremely interesting paper \cite{hh} have presented evidence
that local operators of very high dimensionality contain
information about the size and shape of small\footnote {By small
we do not mean microscopic. The objects we have in mind may be
macroscopic and classical as long as they are much smaller than
the $AdS$ radius.} objects. The resolution of this conflict will
be seen to lie in an ambiguity in what we mean when we say that a
bulk quantity is represented by a certain field theoretic
quantity.

We will begin by very briefly describing the results of Horowitz
and Hubeny. These authors consider objects in $AdS_5\times S_5$
which are much smaller than the AdS radius of curvature and which
are localized at a point on the 5-sphere. In addition they are
also localized near the center of $AdS$ in appropriately chosen
coordinates. Using the gravitational dual theory they find that
the expectation values of certain scalar operators of dimension
$n=l+4$ are of order
\be
\langle \Phi_n \rangle \sim e^{c \rho l} \ee where $\rho$ is the
size of the object in units of the $AdS$ radius and $c$ is a
numerical constant which we will ignore. It follows from eq. (1.1)
that for $l>\rho^{-1}$ the signal is appreciable and that by
examining the $l$ dependence the size of the object can be
determined. This  apparently  contradicts
\cite{pst}\cite{precursors}.

Our conventions for describing $AdS$ space are as follows: Global
coordinates for $AdS$ can be defined so that the metric is given
by
\be
ds^2={R^2 \over (1-r^2)^2}\left\{ (1+r^2)^2 dt^2 -4dr^2 -4r^2
d\Omega^2\right\}\ee where the radial coordinate runs from $0$ to
$1$. Here, $R$ is the radius of curvature of the space and
$d\Omega^2$ represents the metric of a unit sphere. In the case of
$AdS_5$ the sphere is a $3$-sphere.

These coordinates are especially useful when trying to recover
infinite flat space in the limit $R \to \infty$. Indeed the $AdS$
space as defined  above behaves in many respects like a finite
cavity of size $R$ with a reflecting boundary at $r=1$. We will
refer to the metric in eq. (1.2) as cavity coordinates.

Another coordinate system which is particularly useful when
studying the properties of the CFT in flat space is given by
\be
ds^2 ={R^2 \over y^2}{(d\tilde{t}^2 -dx^2 - dy^2)} \ee where
$\tilde{t},x$ labels 4-dimensional Minkowski space and $y$ is the 5th direction
perpendicular to $x$. At time $\tilde{t}=0$, the center of AdS 
can be taken to be the point $x =0$, $y=1$ in these co-ordinates.

The quantum field theory lives on the 4-dimensional boundary whose
metric we take to be either
\be
ds_b^2=dt^2 -d\Omega^2 \ee in the case of cavity coordinates, or
\be
ds^2_b = d\tilde{t}^2 -dx^2 \ee for the flat space representation.

As in ref. \cite{susskind1} we will be thinking of the boundary
quantum field theory in Wilsonian terms. Thus we imagine the
boundary field theory to be defined in terms of a bare set of
degrees of freedom at some very small coordinate length scale
$\delta_0$. The details of the cutoff are not important but an example
to keep in mind is the Hamiltonian lattice cutoff sometimes used to study
QCD. In that case $\delta_0$ would be the spatial lattice
constant. We assume $\delta_0$ is much smaller than any other
length scale we will encounter.

According to the UV/IR correspondence, a connection exists between
the UV regulator length scale $\delta_0$ of the QFT and an IR
cutoff of the bulk theory. The bulk cutoff is implemented by
replacing the $AdS$ boundary at $r=1$ or $y=0$  with the surface
$r=1-\delta_0$ in global $AdS$ or $y=\delta_0$ in the Poincare patch.

There are two different large $N$ limits of the QFT that have two
different purposes. The first is the 't Hooft limit \bea N&\to
&\infty \cr g_s N &=& constant. \eea From the bulk point of view
this is the limit of classical gravity or classical string 
theory in an $AdS$ space of fixed radius in string units;
\be
R=(g_s N)^{1/4} l_s. \ee In this limit the ratio of the size of a
physical object to the $AdS$ radius is fixed.

The limit of interest for analyzing the holographic principle as
defined in \cite{'thooft}\cite{susskind} is a different one
\cite{susskind2}\cite{polchinski}. For this purpose we take \bea
N&\to &\infty \cr g_s  &=& constant. \eea In this limit the $AdS$
radius becomes much larger than the size of any physical object.
This is the limit discussed in \cite{pst} where it was claimed
that objects at $r=0$ or $y=1$ should be represented by Wilson
loops of roughly unit size.

\setcounter{equation}{0}
\section{What it Means to Represent}

One reason for confusion is that different people may mean
different things when they say a certain field theory quantity
represents a corresponding bulk quantity. In order to resolve the
paradox raised by the Horowitz-Hubeny result we need to have a
clear understanding of what it means for a particular set of
observables in the CFT to describe a particular set of
circumstances in the $AdS$ space. 

The ability to find observables (hermitian operators) in the field
theory to represent physical quantities in the bulk theory follows
from the assumption that the Hilbert space of bulk states is the same as
that of the boundary field theory. We argue that $faithfully$
representing a given bulk quantity $\alpha$ by a field theory
quantity $A$ should mean more than just requiring a correspondence
between their expectation values. Ideally we would like the
probability distribution for $\alpha$ and $A$ to be the same. For
example, a faithful representation of a highly classical bulk
quantity such as the size of a macroscopic object should involve a
field theory quantity with very small fluctuation.

Let us suppose that the quantum state of the system determines a
probability distribution $P(\alpha)$ centered at $\alpha_0$ with a
width $\Delta(\alpha)$. Now consider a second state characterized
by a second distribution $P'(\alpha)$ centered at $\alpha'_0$.
These two states are clearly distinguishable if the two
distributions do not overlap. In particular two different
macroscopic classical configurations of $\alpha$ should have
negligible overlap in their probabilities. A minimum condition for
$A$ to faithfully represent $\alpha$ is that two probability
distributions for $A$ will not overlap if the two corresponding
distributions do not overlap for $\alpha$. In other words two
configurations which disagree on the value of $\alpha$ must be
represented by probability distributions in $A$ which are almost
orthogonal. We will regard this to be a minimal requirement for a
faithful representation of a bulk variable by a corresponding
holographic variable.

\setcounter{equation}{0}
\section{The Resolution}
Let us now ask whether the high dimension operators $\Phi_n$ are a
faithful representation  of the size of an object at the center of
$AdS$ space. From what we have said in the previous section the
question comes down to whether or not the probability
distributions for the $\Phi$'s are orthogonal or almost orthogonal
for two classically distinguishable values of the size $\rho$. We
emphasize again that we are working in a Wilsonian framework where
it is assumed that the field theory is defined by a concrete
regularized system.

The $\Phi_n$'s are defined to have vanishing vacuum expectation
values. We must also specify a convention for normalizing them. We
follow the same convention as in \cite{hh}, namely the two point
function $\langle \Phi_n(x) \Phi_n(x')\rangle$ is of order one at
unit coordinate separation. Now consider the width of the
probability distribution for $\Phi_n$, in other words the
fluctuation $\Delta$ in $\Phi_n$:
\be
\Delta^2 = {\langle \Phi_n^2\rangle}.\ee Obviously if the difference
in expectation values of $\Phi_n$ for two distinct configurations is
much less than $\Delta$ then these variables do not faithfully
represent the variables they were intended to describe.

The scalar fields $\Phi_n$ are given by
\be
\Phi_n =TrF^2X^l \ee where $X$ stands for the six fundamental
scalars of maximally supersymmetric $SU(N)$ Super Yang-Mills
Theory. The trace is over the adjoint representation of $SU(N)$
and $X^l$ represents a polynomial of order $l$ in the $X's$. The
dimension of $\Phi_n$ is $n=l+4$. The operators are normal ordered
meaning that their vacuum expectation value has been subtracted
out. They are normalized so that their two point function at unit
coordinate separation is of order one.

Now consider the fluctuation in $\Phi_n$. This is given by the
square root of the connected two point function at vanishing
separation. In the continuum theory this fluctuation will be
divergent. In the Wilsonian cutoff theory the fluctuation will be
of order $\delta_0^{-n}$ which we assume is extremely large
\footnote{This is true for the expectation value in any generic low energy
state since the expectation value will be dominated by high energy
intermediate states.}. Thus unless these operators are somehow further
regulated the fluctuation is divergent.

This is true for any local operator $\Phi$. It means that $\Phi$ can not
faithfully describe anything. A measurement of $\Phi$ gives
completely random results in any state. This point was made
forcefully in a famous paper by Bohr and Rosenfeld in the earliest
days of quantum field theory. According to Bohr and Rosenfeld the
correct observables for a quantum field theory are what we would
today call ``regulated" fields. This entails introducing a
regulator scale $\delta$ chosen to be much larger than the cutoff
scale $\delta_0$. The observables are defined by some form of
smearing or point-splitting of the composite operators $\Phi$.
This will be discussed further in the next section.

The regulated fluctuation in $\Phi_n$ is of order
\be
\Delta_n \sim \delta^{-(l+4)}.\ee This follows from dimensional
analysis and the fact that $\Phi_n$ has mass dimension $n=l+4$.

Evidently from eq. (1.1), the condition that the expectation value
of $\Phi_n$ is larger than the fluctuation is
\be
e^{c \rho l} \geq \delta^{-(l+4)}.\ee Since $\rho$ is defined to be
the size of an object measured in units of $R$, it will vanish in
the limit $R \to \infty$. Thus we find that the inequality is
satisfied only if
\be
\delta \sim 1. \ee

In other words the operators $\Phi_n$ must not only be regulated but
the regulator scale has to be comparable to the coordinate
distance of the object from the $AdS$ boundary. The meaning of
this is clear. For an operator to faithfully represent a property
of a small object near the center of $AdS$, it must be non-local
as described in \cite{pst}\cite{precursors}.

\setcounter{equation}{0}
\section{Regulating $\Phi$ }
Granted that we must regulate the operators $\Phi$, the question
arises as to exactly how to do so. We will begin with an implicit
construction. The problem with the unregulated operators is that
they have large matrix elements connecting two very high energy
states. In calculating their fluctuation most of the contribution
comes from these high energy intermediate states. We can easily
regulate the operators by simply throwing away the matrix elements
between states whose energies differ by more than $\delta^{-1}$.
Equivalently we can integrate the operators over time with a
smooth test function with support over a time interval $\sim
\delta$. By solving the equations of motion we can express the
regulated operator in terms of operators at a fixed time. The
result will be spatially nonlocal over a scale $\delta$. As we
have seen in the previous section, $\delta$ must be $\sim 1$ so
that the regulated operator is nonlocal over the entire boundary
sphere.

On the other hand smearing an operator over time will not change
its expectation value in a time independent configuration. Thus,
for such configurations, the local operator and its nonlocal
counterpart have the same expectation value. This accounts for the
results in \cite{hh}. 
However, for time dependent configurations
such as those described in \cite{pst}\cite{precursors} only
the nonlocal operator faithfully represents the relevant
instantaneous property of a small object near $r=0$. 

The field theory of interest in this paper is a gauge theory in
which all fundamental fields are in the adjoint representation. If
the bare theory is a Hamiltonian lattice gauge theory, then any
operator at a fixed time can be expressed in terms of generalized
Wilson loops in which the Wilson loop is ``decorated" with
insertions of adjoint fields. The regulated operators will be
expressible as linear superpositions of such Wilson loops of size
$\delta$. Horowitz and Hubeny provide important information on how to
decorate the Wilson loop in order to describe particular features of
small objects.

Another important example concerns the ``precursors'' described in
\cite{pst}\cite{precursors}. Suppose that an event takes place near
the center of $AdS$ which results in the emission of a wave
propagating towards the boundary. Bulk causality ensures that all
local supergravity fields evaluated within a neighborhood of the
boundary will retain their original expectation values until the wave
itself arrives at the boundary. Therefore, for a period of time of
order one, all local QFT operators corresponding to the bulk supergravity fields
will retain their original expectation values carrying no information
about the wave. At some time, when the wave arrives at the boundary ($t=0$),
some local operators in the QFT will begin to oscillate. According to
the AdS/CFT correspondence, their expectation value at $t=0$ will
be given by the boundary data of the wave. Furthermore, their expectation value
will be insensitive to the $R \rightarrow \infty$ limit and
proportional to the amplitute of the wave. Thus, in regulating
the local operators at $t=0$ so as to keep the signal bigger than
their fluctuations, we only need to introduce a cutoff of order the width of the
wave pulse. Now to find the non-local ``precursors'' describing the
wave at an earlier time, when say the wave is at co-ordinate distance
$\delta$ from the boudary, we use the equations of motion to express
the regulated local operators at $t=0$ in terms of operators at
$t=-\delta$. Then, the ``precursors'' will be spatially
non-local over a scale $\delta$. The results of \cite{precursors}
suggest that the resulting non-local operator will involve
superpositions of Wilson loops of size $\delta$.    

A point worth mentioning involves the possibility of constructing
operators with small fluctuation by spatially averaging $\Phi$
over $\Omega$. It is not hard to see that this diminishes its
fluctuation by a factor $\delta^{3/2}$. This would have no
important effect on our conclusion.

Finally we want to emphasize that expectation values are not the
observables of a system. The observables representing the results
of measurements have uncertainties. A correct representation of a
variable should not only represent its expectation value but also
its entire probability distribution. The wild fluctuation of local
fields makes them bad representations of weakly fluctuating
positions and sizes of macroscopic objects.




\section{Acknowledgements}
We would like to acknowledge G. Horowitz and S. Thomas for useful
discussions. The work of V. Hubeny was supported in part by NSF grants
PHY-9507065 and PHY-0070895. The work of L. Susskind and N. Toumbas
was supported in part by NSF grant PHY-9870115.




\end{document}